\documentclass[final,3p,times]{elsarticle}

\usepackage{amssymb}
\usepackage{amsmath}

\journal{Journal of Subatomic Particles and Cosmology}

\begin{document}

\begin{frontmatter}



\title{Probing collective behaviour of Heavy Quarks
through $p_T$-differential radial flow $v_0(p_T)$}


\author[aaa,bbb]{Salvatore Plumari}
\author[aaa,bbb]{Maria Lucia Sambataro}
\author[ccc]{Santosh K. Das}
\author[aaa,bbb]{Vincenzo Greco}

\affiliation[aaa]{organization={Department of Physics and Astronomy “Ettore Majorana”, University of Catania},
             addressline={Via S. Sofia 64},
             city={Catania},
             postcode={I-95123},
             state={Catania},
             country={Italy}}

 \affiliation[bbb]{organization={INFN-Laboratori Nazionali del Sud},
             addressline={Via S. Sofia 62},
             city={Catania},
             postcode={I-95123},
             state={Catania},
             country={Italy}}

 \affiliation[ccc]{organization={School of Physical Sciences, Indian Institute of Technology Goa},
             city={Ponda},
             postcode={403401},
             state={Goa},
             country={India}}

\begin{abstract}
We discuss the $p_T$-differential radial flow $v_0(p_T)$ of charmed hadrons within a Langevin dynamics coupled to relativistic Boltzmann transport approach in an event-by-event basis. We propose heavy flavour $v_0(p_T)$ as a novel observable to probe the strength of the interaction of heavy quarks with the expanding Quark-Gluon Plasma. By comparing different temperature dependence for the spatial diffusion coefficient $D_s(T)$ we show that the $v_0(p_T)$ keep a strong sensitivity to the heavy-quark transport coefficients at intermediate $p_T$. At low $p_T$, the observable is also sensitive to hadronization where we observe a larger $v_0(p_T)$ for $\Lambda_c$ baryons than for $D$ mesons. 
\end{abstract}




\end{frontmatter}


\section{Introduction}
Ultra-relativistic heavy-ion collisions at RHIC and at the LHC energies 
provide a laboratory tool to study strongly interacting matter at temperatures and energy densities above the QCD crossover. The created matter shortly after the collision is expected to undergo a rapid evolution from an initially far-from-equilibrium state toward a deconfined quark-gluon plasma (QGP) and at most thermalized, followed by expansion, cooling, hadronization and subsequent hadronic evolution. The produced QCD matter behaves as a strongly coupled medium, whose collective expansion is very well described by relativistic viscous hydrodynamics.
The most widely used observables to get information about the transport coefficient of the matter created are the anisotropic flow coefficients $v_n(p_T)$. These observables are generated by the conversion of the initial anisotropies in cordinate space into final momentum anisotropies. In contrast, the isotropic component of the expansion, usually referred to as radial flow, has been studied through the shape of transverse-momentum spectra or through the mean transverse momentum $\langle p_T\rangle$. \\
The recently introduced observable in the light sector the radial flow correlator $v_0(p_T)$ fills this gap. It was first proposed in Ref. \cite{Schenke:2020uqq} as the isotropic analogue of the usual differential flow coefficients $v_n(p_T)$, and further developed in hydrodynamic studies \cite{Parida:2024ckk}. The observable measures the correlation between the fluctuation of the particle yield in a given transverse momentum $p_T$ bin and the event-by-event fluctuation of the mean transverse momentum. In other words, $v_0(p_T)$ quantifies how the differential spectrum changes when an event has a larger or smaller radial flow than the event average. Recently, the ATLAS and ALICE collaboration have measured this observable in Pb-Pb collisions at $\sqrt{s_{\rm NN}}=5.02 \,  TeV$ \cite{ATLAS:2025ztg,ALICE:2025iud}. Further investigations  have shown that $v_0(p_T)$ is sensitive to bulk properties of the medium, including the bulk viscosity and the QCD equation of state \cite{Parida:2024ckk, ATLAS:2025ztg}. In ref. \cite{Sambataro:2025pop} this observable was extended for the first time to the heavy-quark sector. This is a physically important step because heavy quarks are qualitatively different from light quarks and gluons. Charm and bottom quarks are produced in the initial hard scatterings, on time scales of order $1/m_Q$, and their number is approximately conserved during the QGP evolution. They therefore propagate through the full space-time evolution of the medium and can keep memory of the microscopic interaction with the QGP. The main observables of heavy-flavor, such as the nuclear modification factor $R_{\rm AA}(p_T)$ and the elliptic flow $v_2(p_T)$, are already used to constrain the interaction of HQs with the the medium and extract information about the spatial diffusion coefficient of the heavy-quark $D_s(T)$ 
\cite{Rapp:2018qla, Dong:2019unq,Moore:2004tg,Gossiaux:2008jv,Alberico:2011zy,Das:2013kea,Das:2015ana,Song:2015ykw,Cao:2018ews,Katz:2019qwv,Plumari:2019hzp,Zhao:2024oma}. 
The $v_0(p_T)$ for heavy quarks measures how efficiently the event-by-event fluctuations of the expanding bulk are transferred to a Brownian motion of the HQs and it gives a direct access to the strength of the HQs medium coupling. The important question addressed by the work is whether this observable can discriminate between weakly and strongly coupled heavy-quark transport scenarios and whether it can also provide information on heavy-flavour hadronization.
\section{Transport approach}
The calculation is performed within an event-by-event transport framework in which the evolution of the QGP and the charm quarks are described by different dynamical equations. The light partons are described by a relativistic Boltzmann transport equation at fixed shear-viscosity-to-entropy-density ratio $\eta/s$ \cite{Plumari:2012ep,Scardina:2012mik,Scardina:2014gxa,Scardina:2017ipo,Sambataro:2020pge,Plumari:2019gwq,Plumari:2019hzp,Sambataro:2022sns,Sambataro:2023tlv,Sambataro:2025obe,Nugara:2025ueb,Nugara:2024net}. The evolution of the distribution function $f_j(x,p)$, is given by
\begin{equation}
p_j^\mu \partial_\mu f_j(x,p)=C[f_j,f_i](x,p), \, \, with \, \, j=g,q
\end{equation}
where $C[f_j,f_i]$ contains only elastic scattering processes. 
The heavy quarks are treated as Brownian particles propagating through the expanding medium. Their dynamics is described by the relativistic Langevin equations
\begin{equation}
dx_j=\frac{p_j}{E}\,dt,
\qquad
dp_j=-\Gamma p_j\,dt+\sqrt{dt}\,C_{jk}\rho_k ,
\end{equation}
where $\Gamma$ is the drag coefficient and $C_{jk}$ is related to the diffusion coefficient of the momentum-space and for isotropic diffusion tensor we have $C_{jk}=\sqrt{2D_p(E)}\,\delta_{jk}$.
The fluctuation-dissipation theorem then relates the momentum-space diffusion coefficient to the drag coefficient through
$D_p(p)=A(p)E(p)T$.
\begin{figure}[]\centering
\includegraphics[width=0.40\linewidth]{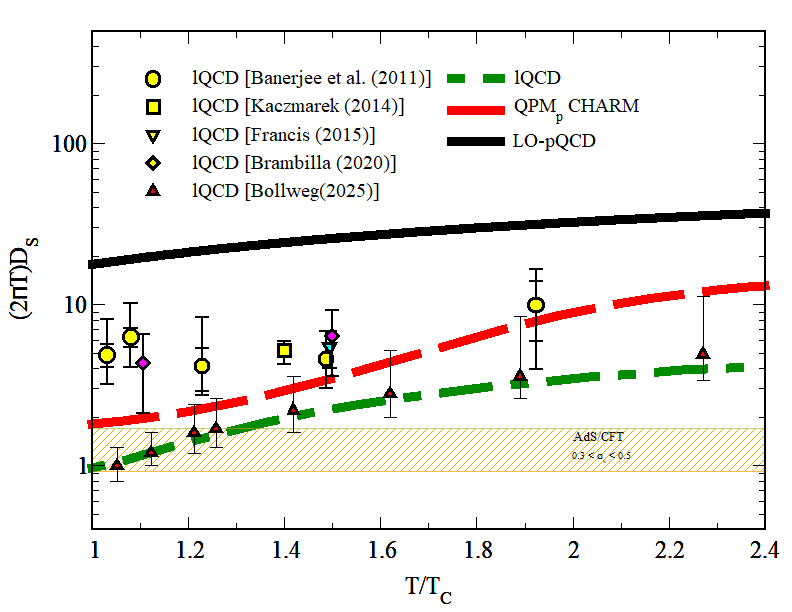}
\hspace{0.5 cm}
\includegraphics[width=0.37\linewidth]{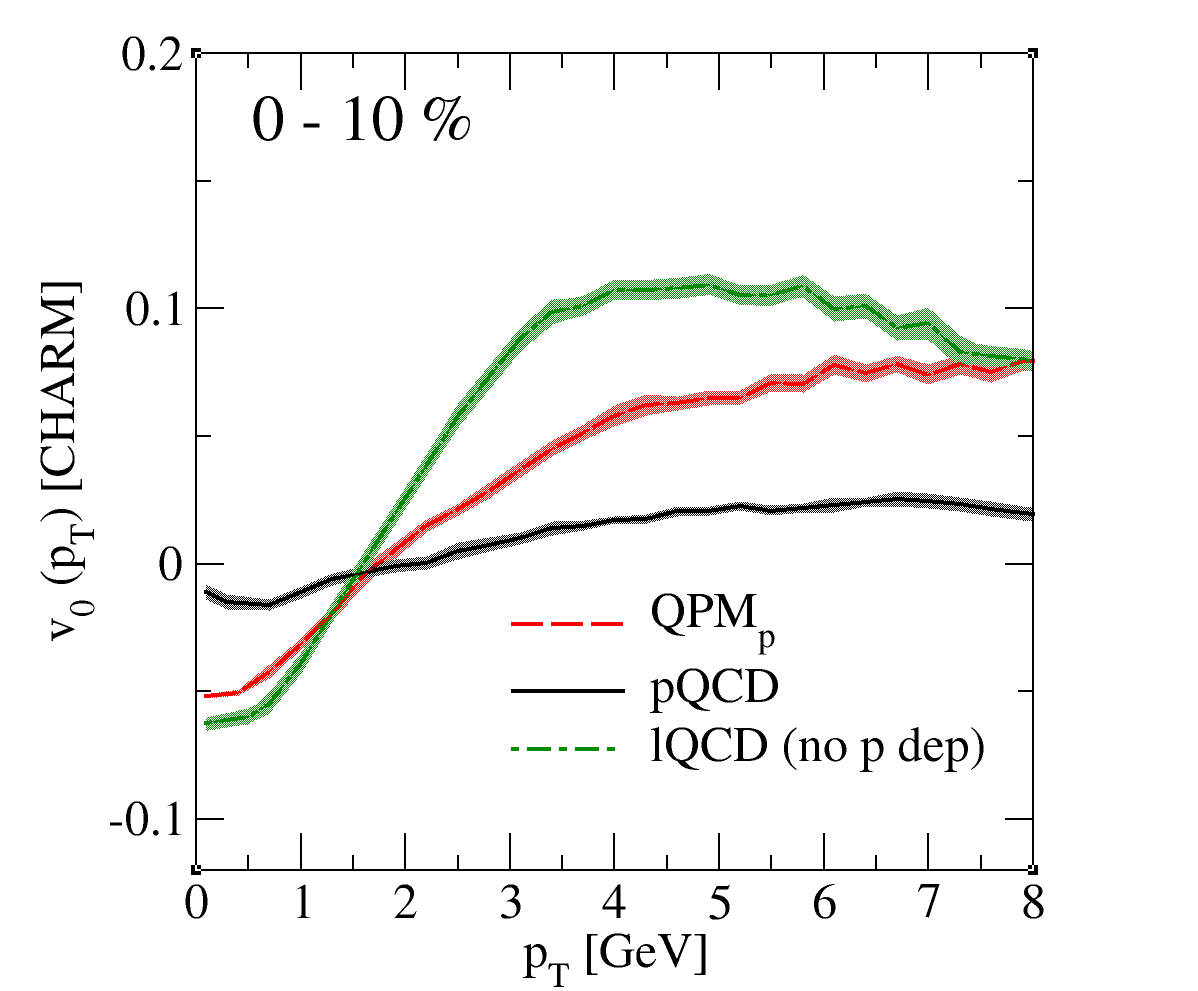}
\caption{Left panel: Spatial diffusion coefficient $2\pi T D_s$ used in this work and compared to lQCD data \cite{Altenkort:2023eav,Brambilla:2020siz,Francis:2015daa,Kaczmarek:2014jga,Banerjee:2011ra,Altenkort:2023oms}. Right panel: $v_0(p_T)$ of charm quarks for Pb-Pb collisions and $0-10\%$ centrality cut same different colors are for different interactions.  }\label{fig1}
\end{figure}
In the static limit, the drag coefficient can be related to the spatial diffusion coefficient $D_s$.\\ 
We present a study of the $p_T$-differential radial flow $v_0(p_T)$ of charm quarks in the QGP. Following the definition of $v_0(p_T)$ in~\cite{Schenke:2020uqq} we have:
\begin{equation}
  v_0(p_T) \equiv
  \frac{\langle \delta N(p_T)\,\delta p_T\rangle}
       {N_0(p_T)\,\sigma_{p_T}},
\end{equation}
which quantifies the event-by-event correlation between a fluctuation
$\delta N(p_T)$ of the single-particle spectrum at transverse momentum
$p_T$ and the fluctuation $\delta p_T$ of the mean transverse momentum
per particle.\\
In order to investigate the sensitivity of $v_0(p_T)$ to the heavy quark–bulk interaction, we explore three different scenarios going from a weakly coupled regime, as described by perturbative QCD (pQCD), to a strongly interacting medium, as suggested by recent unquenched lattice QCD (lQCD) results \cite{Altenkort:2023eav, Altenkort:2023oms} as shown in the left panel of Fig.~\ref{fig1}. The first is a weak-coupling setup based on leading-order pQCD drag and diffusion coefficients with a running coupling \cite{Kaczmarek:2005ui}, (black solid line) . The second is a strong-coupling scenario with momentum-independent $D_s$, fitted to recent unquenched lQCD results \cite{Altenkort:2023oms,Altenkort:2023eav} (green dot-dashed line).  The third scenario employs the recently developed $QPM_p$ (red dashed line), where momentum-dependent parton masses are included \cite{Sambataro:2024mkr,Sambataro:2025obe}. 
In the right panel of Fig.~\ref{fig1} we show the $v_0(p_T)$ of charm quarks before hadronization in Pb-Pb collisions at $\sqrt{s_{\rm NN}}=5.02 \, TeV $. The three curves correspond to the three interaction scenarios discussed above: pQCD, QPMp, and the strong lattice-QCD-inspired case without momentum dependence.
At low $p_T$, $v_0(p_T)$ is negative and around a transverse momentum of the order of the charm quark mass, the observable crosses zero. At larger $p_T$, it becomes positive, reflecting the enhancement of the spectrum in events with stronger radial flow.
We observe that the magnitude of $v_0(p_T)$ is sensitive to the microscopic charm-medium coupling. The $v_0(p_T)$ for pQCD interaction is very small and almost flat, indicating that a weakly coupled charm quark does not efficiently convert the information of the radial fluctuations of the QGP bulk. The QPMp curve is much larger, roughly a factor of three above the pQCD result at intermediate $p_T$. However, the lQCD inspired scenario produces an even stronger effect. This hierarchy comes from the fact that the smaller the spatial diffusion coefficient $D_s$, the stronger the coupling between the heavy quark and the bulk medium, and the larger the transfer of event-by-event fluctuations to the charm spectrum and suggests that this observable can be used to obtain information about the HQ-bulk interaction.
\begin{figure}[]\centering
\includegraphics[width=0.40\linewidth]{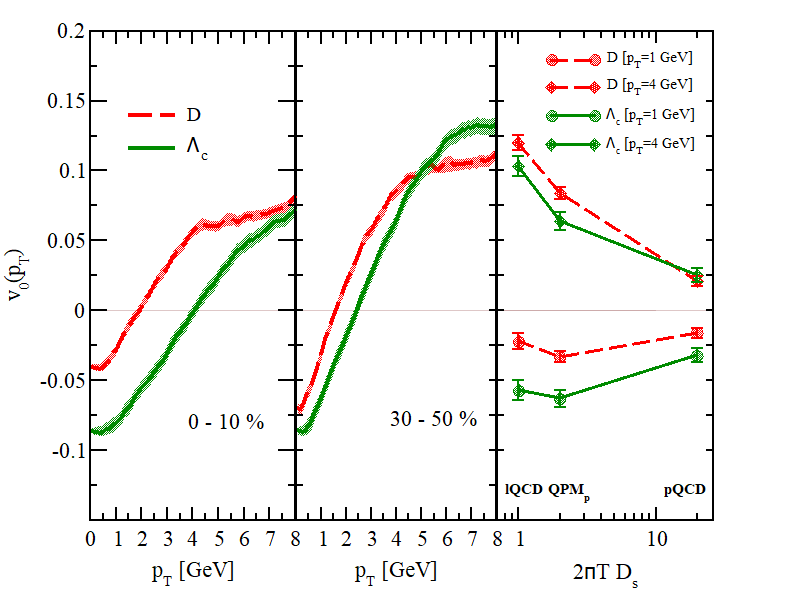}
\caption{$v_0(p_T)$ at $0- 10 \, \%$ (left) and $30-50 \, \%$ (middle) for D (red lines) and $\Lambda_c$ (green lines) in $QPM_p$ from coalescence plus fragmentation. (right) D (red line) and $\Lambda_c$ (green line) $v_0(p_T)$ at $30 -50 \, \%$ for $p_T=1 \, GeV$ (circles) and $p_T=4 \, GeV$ (diamonds) as function of $2\pi T D_s$.}\label{fig2}
\end{figure}
In Fig.~\ref{fig2} we show the $v_0(p_T)$ after hadronization for both $D$ (red lines) and $\Lambda_c$ (green lines) in the QPMp scenario and using the hadronization approach by coalescence plus fragmentation. In left and middle panels we compare $v_0(p_T)$ of mesons and baryons for $0\text{-}10\%$ and $30\text{-}50\%$ centrality classes and in the right panel we show the dependence of $v_0$ on $2\pi T D_s$ at fixed transverse momenta, $p_T=1 \, GeV$ and $p_T=4\, GeV$. 
The $v_0(p_T)$ of $\Lambda_c$ crosses zero at a larger transverse momentum than the $D$, similarly to that observed for pions and protons. 
At low $p_T$ and around $p_T\simeq 1\, GeV$ the difference can be a factor of two and this makes the comparison between $v_0(D)$ and $v_0(\Lambda_c)$ a potential fingerprint of hadronization.
In the right panel of Fig.~\ref{fig2} we show that at intermediate $p_T$ 
the value of $v_0(p_T)$ shows a strong dependence on $2\pi T D_s$ and therefore sensitive to the heavy-quark interaction where stronger interaction leads to a larger response of the heavy quark to the fluctuating bulk medium. 
Conversely, at lower $p_T$ the dependence on $D_s$ is much weaker, while the difference between $D$ and $\Lambda_c$ is large. Therefore, the low $p_T$ region is more sensitive to hadronization, whereas the intermediate $p_T$ region is more affected by the heavy-quark interaction.
\section{Conclusions}
We have introduced $v_0(p_T)$ as a new heavy-flavor observable that is sensitive to the interaction between charm quarks and the QGP. The charm $v_0(p_T)$ clearly discriminates between weak pQCD-like interactions and stronger non-perturbative scenarios such as QPMp or lQCD inspired interactions. After hadronization, the observable remains sensitive to the transport coefficient at intermediate $p_T$, while at low $p_T$ it becomes a promising observable sensitive to hadronization, predicting a split between $D$ and $\Lambda_c$. 
%
%



\end{document}